\documentclass[prd,twocolumn,superscriptaddress,showpacs,preprintnumbers,nofootinbib,APS]{revtex4}
\usepackage{color}
\usepackage{psfrag}
\usepackage{dcolumn}
\usepackage{bm}
\usepackage[latin1]{inputenc}
\usepackage[spanish,english]{babel}
\usepackage{amsfonts}
\usepackage{amssymb,epsf}
\usepackage{graphicx}
\usepackage{epstopdf}
\usepackage{epsfig}
\usepackage{latexsym}
\usepackage{epsf}

\begin{document}

\title{Range of novel black hole phase transitions via massive gravity:\\
Triple points and  \textit{N}-fold  reentrant phase transitions }

\author{A. Dehghani}
\affiliation{{\footnotesize Physics Department and Biruni
Observatory, College of Sciences, Shiraz University, Shiraz 71454,
Iran}}
\author{S. H. Hendi}
\affiliation{{\footnotesize Physics Department and Biruni
Observatory, College of Sciences, Shiraz University, Shiraz 71454,
Iran}}
\author{R. B. Mann}
\affiliation{{\footnotesize Department of Physics and Astronomy,
University of Waterloo, 200 University Avenue West, Waterloo,
Ontario, Canada N2L 3G1}}

\begin{abstract}
Massive gravities in anti-de Sitter spacetime can be viewed as effective dual field theories of different
phases of condensed matter systems with broken translational symmetry such as solids, (perfect) fluids, and liquid crystals. Motivated by this fact, we explore the black hole chemistry (BHC) of these theories and find a new range of novel phase transitions close to realistic ones in ordinary physical systems. We find that the equation of state of topological black holes (TBHs) at their inflection point(s) in $d$-dimensional spacetime
reduces to a polynomial equation of degree $(d-4)$, which yields up to $n=(d-4)$ critical points. As a result, for (neutral) TBHs, we observe triple-point phenomena with the associated first-order phase transitions (in $d \ge 7$), and a new phenomenon we call an $N$-fold reentrant phase transition, in which several ($N$) regions of thermodynamic phase space exhibit distinct reentrant phase transitions, with associated
virtual triple points and zeroth-order phase transitions (in $d \ge 8$), as well as Van der Waals transitions (in $d \ge 5$) and reentrant (in $d \ge 6$) behavior. We conclude that BHC in higher-dimensional massive gravity
is very likely to offer further new surprises.
\end{abstract}

 \maketitle



\section{Introduction}

From a classical field theory
perspective, dRGT massive gravity \cite{deRham2010Gabadadze,dRGT}
is a consistent extension of general relativity with an explicit
mass term. By giving the graviton a mass, massive gravity could
provide a possible explanation for the accelerated expansion of
the universe without the requirement of dark energy
\cite{MassiveCosmologies2011,Akrami2013Koivisto,Akrami2015Hassan}.
Moreover it has recently been shown that massive spin$-2$
particles can explain the current observations related to dark
matter \cite{Schmidt-May2016JCAP}. Furthermore, the new
observations from LIGO \cite{LIGO2017} imply that the graviton
mass is bounded to ${m_g} \le 7.7 \times {10^{ - 23}}eV/{c^2}$,
and so such an assumption remains empirically viable, since models
of massive gravity typically yield the bound ${m_g} \le {10^{ -
30}-10^{ - 33}}eV/{c^2}$, whose observable effects are out of
reach of LIGO \cite{deRhamREVIEW2014}.

Perhaps the most challenging problem in building a massive
gravitational field theory is the appearance of ghosts.  It has
long been known that  some specific models of massive gravity
suffer from ghost instabilities (so-called Boulware-Deser (BD)
ghosts) \cite{BDghost1972}. However dRGT massive gravity has been
shown to be ghost-free in the decoupling limit to all orders in
the nonlinearities; furthermore, away from the decoupling limit,
the possibility of ghost fields is excluded at least up to and
including quadratic order in the nonlinearities \cite{dRGT}. It
was later  shown that any pathological BD ghost is eliminated at
the full nonlinear level due to the Hamiltonian constraint and the
existence of a nontrivial secondary constraint, which yields
enough conditions to remove such ghosts \cite{HassanRosen2012PRL}.
This ghost analysis  generalizes both to
 arbitrary massive couplings
\cite{HassanRosenSchmidt-May2012JHEP} and higher dimensions  \cite{TQDO2016a,TQDO2016b}.

Beyond these achievements, dRGT massive gravity in AdS with a
singular, degenerate reference metric is proving useful in
building holographic models for normal conductors that are close
to realistic ones, with  finite DC conductivity
\cite{Vegh2013,BlakeTong2013,Alberte2016}. Recent developments
show that they can be regarded as dual to effective field theories
of different phases of matter, particularly  homogeneous and
isotropic condensed matter systems with broken translational
invariance
\cite{Alberte2016SolidHolography,Alberte2018HolographicPhonons,Alberte2018BHelasticity}.

For these reasons we study here the chemistry of AdS
black holes (BHC) in this context, where thermodynamic pressure
$P=-\Lambda/8\pi$, with $\Lambda <0$ is the cosmological constant
\cite{CQG2017Review}. We concentrate on higher dimensional massive
gravity as an alternative candidate to Einstein's general
relativity, investigate the extended phase space thermodynamics of
the AdS BH solutions in detail, and bring some new perspectives on
BH thermodynamics with massive gravitons. The phase structure of
this class of theories has been yet not considered in higher
dimensions at all orders.  Our purpose is to discover  which
properties of black holes are universal and which  show a
dependence on the spacetime dimension. In so doing we find
a new range of critical phenomena that may also be present in
nature. Specifically, we find that Topological Black Holes (TBHs)
with massive gravitons can mimic the critical behavior of
everyday substances in nature without the inclusion of any extra
or unusual matter fields in the gravitational action. We find
that, besides van der Waals (vdW)
\cite{PVmassive2015,TBH2017MassiveGravity} and reentrant
\cite{ReentrantMassive2017}  phase behaviors, in massive gravity a
gravitational triple point can emerge for AdS BHs with various
horizon topologies in spacetime dimensions with $d \ge 7$.

Even more remarkably, in $d \ge 8$ we discover a novel and
more complex phenomenon: $N$-fold reentrant phase transitions
(RPTs), for all types of TBHs. By ``$N$-fold RPTs"
we mean that a number of RPTs are present in distinct
regions of thermodynamic phase space, i.e., for regions
$ P \in (P_{Tr_{i}},P_{Z_{i}}) $
and $ T \in (T_{Tr_{i}},T_{Z_{i}})$ the associated $\rm{RPT}_{i}$
occurs. This is in contrast to the situation in Lovelock
gravity (first seen in Ref. \cite{Frassino2014}), in which hyperbolic black
holes exhibit double (or multiple) RPTs (with the associated
sequence of LBH $\to$ SBH $\to$ LBH $\to$ SBH ... phase
transitions, where SBH denotes small black hole, LBH
denotes large black hole and IBH denotes intermediate
black hole) as the temperature decreases along a single line
in phase space; this critical behavior is similar to observed
double reentrant transitions for smectic and nematic phases
of liquid crystals \cite{DoubleReentrance1980,DoubleReentrance1988,DoubleReentrance1989,DoubleReentrance2003}, which schematically have the
form $A \to B \to A \to B$. By contrast, $N$-fold RPTs refer to
the phenomenon of RPTs (multiple or not) occurring along
more than one line of temperature in phase space.

This situation, in some aspects, resembles that of a
(experimentally confirmed) scenario for liquid crystals
\cite{MultipleRPTs1986,MultipleRPTs1988Review,MultipleRPTs2002Book}, but in a slightly different way. In liquid crystals,
reentrance is encountered as the temperature is lowered
monotonically, with other thermodynamic quantities kept
fixed. For massive gravity TBHs we find that several RPTs
occur at various locations in phase space. From the
experimental point of view, since it is not easy to obtain
RPTs from any theory \cite{MultipleRPTs1988Review}, we believe that these kinds of
exact and simple relations in modified gravity could
possibly shed some light on the microscopic structure of
this strange phenomenon and establish a new link between
BH physics and the realm of statistical mechanics of manybody
systems.

$N$-fold RPTs are thus a generic feature of higher dimensional
TBHs in massive gravity. We explicitly show
that the analogue of triple points in SBH/IBH/LBH phase
transitions and virtual triple points in $N$-fold RPTs can be
obtained by adding the fifth-, sixth-, and higher-order
graviton self-interactions besides the first four terms that
usually appear in the literature. Although we do not find
evidence for any quadruple critical points  \cite{NarayananKumar:1994}, their
existence is not ruled out; whether or not such phenomena
exist for BHs (in massive gravity or elsewhere) remains an
open question.


\section{Action, field equations and AdS black holes}
The bulk action for massive gravity on the $d$-dimensional
background manifold $\cal M$ in the presence of negative
cosmological constant can be written as \cite{deRham2010Gabadadze, dRGT}
\begin{equation} \label{bulk action}
{{\cal I}_b} =  - \frac{1}{{16\pi {G_d}}}\int_{\cal M}
{{d^d}x\sqrt { - g} \Big[ {R - 2\Lambda  + m_g^2\sum\limits_{i =
1}^{d - 2} {{c_i}{{\cal U}_i}(g,f)} } \Big]},
\end{equation}
where the overall minus ensures that the semi-classical
partition function yields consistent results for the
thermodynamic quantities. Here $m_g$ is the
graviton mass, $c_{i}$'s are arbitrary massive couplings, and
\begin{equation}
{{\cal U}_i} = \sum\limits_{y = 1}^i {{{( - 1)}^{y + 1}} \frac{{(i
- 1)!}}{{(i - y)!}}} {{\cal U}_{i - y}}[{{\cal K}^y}],
\end{equation}
where ${\cal U}_{i - y}=1$ if $i=y$. The massive graviton self-interactions
${{\cal U}_i}$ are constructed from a $d \times d$ matrix ${\cal
K}_{\,\,\,\,\,\nu }^\mu$, which is posited to have the following
explicit form
\begin{equation}
{\cal K}_{\,\,\,\,\,\nu }^\mu  = (\sqrt {\cal K}
)_{\,\,\,\,\lambda }^\mu (\sqrt {\cal K} )_{\,\,\,\,\nu }^\lambda
= \sqrt {{g^{\mu \lambda }}{f_{\lambda \nu}}}.
\end{equation}
${g_{\mu \nu }}$ is the dynamical (physical) metric, and
${f_{\lambda \nu}}$ is the auxiliary reference metric, needed
to define the mass term for gravitons.

We seek here a  holographic system (AdS BH) that mimics the
physics of solids, liquids, (perfect) fluids, and especially
liquid crystals, i.e., a system with broken (spatial)
translational symmetry as a key ingredient \cite{Science2011}. To
this end, following
\cite{Alberte2016SolidHolography,Alberte2018BHelasticity,Alberte2018HolographicPhonons},
we consider   a subclass of dRGT massive gravity
\cite{Vegh2013,Alberte2013,BlakeTong2013}, having a dynamical
(physical) metric $g_{\lambda \nu}$ and a degenerate (nonphysical)
reference metric ${f_{\lambda
\nu}}={\partial _\lambda}{\phi ^a}{\partial_\nu }{\phi ^b}{f_{ab}(\phi)}$, in the configuration
space of scalar St\"uckelberg fields $\phi^a$ ($a=1,2,...,d$), where the spatial
inhomogeneities are substituted with the graviton mass terms
\cite{Vegh2013}. This is equivalent to working
with $(d-2)$ St\"uckelberg fields for restoring translational symmetry in spatial directions: there is a gravitational
sector with massless gravitons, encoding ${d(d-3)}/2$
physical modes, and a scalar sector
with $(d-2)$ St\"uckelberg fields minimally coupled to gravity \cite{Alberte2013} that encode $(d-2)$ physical degrees of freedom; in all there are $({d(d-1)-4})/2$ degrees of freedom.

By gauge fixing (e.g.,
working in the unitary gauge ${\phi ^a} = \delta _\mu ^a{x^\mu
}$), the general covariance is preserved in the $(t,r)$
coordinates and breaks in the other spatial coordinates
($x_{1},x_{2},...,x_{d-2}$). Consequently, the dual gauge theory
on the AdS boundary will have a conserved energy without conserved
momentum currents
\cite{Vegh2013,BlakeTong2013}. As mentioned, we are looking for a holographic system (AdS BH) that
mimics the physics of solids, liquids, (perfect) fluids, and
especially liquid crystals, i.e., a system with broken
(spatial) translational symmetry as a key ingredient. To do so, we need a system of massless scalar fields
interacting with pure (Einstein) gravity that can be separated
into different phases of physical matter \cite{Alberte2016SolidHolography,Alberte2018BHelasticity,Alberte2018HolographicPhonons}. This
theory can be reformulated in an equivalent way via dRGT
massive gravity with a singular (degenerate) reference
metric, as we shall assume.

We employ the static ansatz
\begin{equation} \label{dynamical metric}
d{s^2} =  - V(r)d{t^2} +V(r)^{-1}{d{r^2}} +
{r^2}{h_{ij}}d{x_i}d{x_j}
\end{equation}
for the dynamical metric ${g_{\mu \nu }}$. The degenerate
(spatial) background \cite{Vegh2013,Cai2015} is chosen
for the reference metric as ${f_{\mu \nu
}} = diag\left({0,0,c_0^2{h_{ij}}} \right)$ where $c_0$ is
positive constant, with
${h_{ij}}d{x_i}d{x_j} = dx_1^2 + \frac{{{{\sin }^2}(\sqrt k
{x_1})}}{k}\sum\nolimits_{i = 2}^{d - 2} {dx_i^2}
\prod\nolimits_{j - 2}^{i - 1} {{{\sin }^2}{x_j}}$
representing  spherical ($k=1$), planar ($k=0$), and hyperbolic ($k=-1$)
horizon geometries of constant curvature $d_{1} d_{2} k$ and volume
$\omega _{d_{2}}^{(k)}$ (in what follows we will use the
convention $d_{n}=d-n$).  With appropriate identifications these become
compact surfaces of higher genus, yielding TBHs \cite{Mann:1997iz}. The interaction
terms ${\cal U}_i$ are   ${{\cal U}_i} = {\left(
{{c_0}/r} \right)^i}\prod\nolimits_{j = 2}^{i + 1} {{d_j}}$;
there exist $(d-2)$ potential terms in a
$d$-dimensional spacetime.

Varying the bulk action (\ref{bulk action}), including the
Gibbons-Hawking surface term (${\cal I}_{GH}$), with respect to
the dynamical metric ($g_{\mu \nu }$) yields
\begin{equation} \label{field equations}
{G_{\mu \nu }} + \Lambda {g_{\mu \nu }} + m_g^2{{\cal X}_{\mu \nu
}} = 0,
\end{equation}
where
\begin{equation}
{{{\cal X}_{\mu \nu }} =  - \sum\limits_{i = 1}^{d - 2}
{\frac{{{c_i}}}{2}\left[ {{{\cal U}_i}{g_{\mu \nu }} +
\sum\limits_{y = 1}^i {{{( - 1)}^y}\frac{{i!}}{{(i - y)!}}{{\cal
U}_{i - y}}{\cal K}_{\mu \nu }^y} } \right]} }.
\end{equation}
The above gravitational field equations can be analytically solved
using metric ansatz (\ref{dynamical metric}), and, the metric
function $V(r)$ is obtained as
\begin{equation} \label{metric function}
V(r) = k + \frac{{{r^2}}}{{{\ell ^2}}} - \frac{m}{{{r^{{d_3}}}}} +
m_g^2\sum\limits_{i = 1}^{d - 2} {\Big(
{\frac{{c_0^i{c_i}}}{{{d_2}{r^{i - 2}}}}\prod\limits_{j = 2}^i
{{d_j}} } \Big)},
\end{equation}
where  the Arnowitt-Deser-Misner (ADM) mass of the black hole is \cite{Cai2015}
\begin{equation} \label{ADMmass}
M = \frac{{{d_2}{\omega _{{d_2}}^{(k)}}}}{{16\pi }}m ,
\end{equation}
 where for up to four interaction
potentials ($i=1,2,3,4$)
 \begin{eqnarray}
 m &=& kr_ + ^{{d_3}} + \frac{{r_ +
^{{d_1}}}}{{{\ell ^2}}} \\
&&\quad + m_g^2r_ + ^{{d_3}}\left(
{\frac{{{c_0}{c_1}}}{{{d_2}}}{r_ + } + c_0^2{c_2} +
\frac{{{d_3}c_0^3{c_3}}}{{{r_ + }}} +
\frac{{{d_3}{d_4}c_0^4{c_4}}}{{r_ + ^2}}} \right),\nonumber
\end{eqnarray}
with $V(r_{+})=0$;  extension to higher order potentials is
straightforward. 

Note that although the $\Lambda$-term of the metric function (\ref{metric function})
is dominant for large $r$ and the curvature tensor approaches
that of pure AdS spacetime, the asymptotic symmetry
group is not necessarily that of pure AdS. For example
charged black hole solutions with a degenerate reference
metric \cite{Cai2015} have the same form as Eq. (\ref{metric function}) but break the
global symmetries of AdS. To our knowledge there has
been no thorough analysis in the literature regarding the
asymptotic symmetries of solutions in massive gravity with
a negative cosmological constant, and we shall not consider
a full analysis of the asymptotic behavior of our solutions
here.

The Hawking temperature of the BH spacetime can be
obtained by employing the Euclidean formalism. By the analytic
continuation of the Lorentzian metric (\ref{dynamical metric}) to
Euclidean signature and requiring the regularity condition near
the horizon, we obtain
\begin{eqnarray} \label{temperature}
T&=&{\beta ^{ - 1}}  = {\left. {\frac{{V'(r)}}{{4\pi }}} \right|_{r = {r_ + }}} \nonumber \\
&= &\frac{1}{{4\pi {r_ + }}}\Bigg[ {{d_3}k + {d_1}\frac{{r_ +
^2}}{{{\ell ^2}}}+ m_g^2\sum\limits_{i = 1}^{d - 2} {\Big(
{\frac{{c_0^i{c_i}}}{{r_ + ^{i - 2}}}\prod\limits_{j = 3}^{i + 1}
{{d_j}} } \Big)} } \Bigg]
\end{eqnarray}
for the Hawking temperature.


\section{Euclidean Action and Free Energy}
We assume the
gravitational partition function of the massive AdS BH
could be defined by a Euclidean path integral over a dynamical
metric (tensor field $g_{\mu \nu}$) as
\begin{equation} \label{partition function}
{\cal Z} = \int {{\cal D}[g]} {e^{ - {{\cal I}_E}[g]}} \simeq {e^{
- {{\cal I}_{on - shell}}}}
\end{equation}
whose  most dominant contribution
originates from substituting the classical solutions of the
action, i.e. the so-called on-shell action, by applying the saddle
point approximation. The on-shell action can be evaluated using
Hawking-Witten prescription (the so-called subtraction method)
\cite{HawkingPage1983,Witten1998b}, and the divergence in the
partition function will be canceled. That leads to $F  \equiv \Delta F={\beta ^{ - 1}}\left( {{{\cal I}_{BH}} - {{\cal
I}_{AdS}}} \right)$ for the free energy
difference, in which the zero point energy (ZPE) of the boundary gauge theory
is eliminated due to our renormalization method.  We choose the
thermal AdS background in massive gravity as the ground state with
the period $\beta_{0}$, which is different from the period of
massive AdS BH, $\beta$. Note that this background solution is given
by setting $m=0$ in (\ref{metric function}), which is not 
pure AdS;  it will later become
evident that this is the appropriate background.
As usual in massive gravity,
cosmological and black hole solutions are modified at long
distances relative to their counterparts in Einstein gravity.  For
a sufficiently  tiny graviton mass $m_g$, the black hole solution
tends to the Schwarzschild-AdS case at short distances.

In deriving the
on-shell action, we have made use of some necessary ingredients,
which we  briefly explain.
The Ricci scalar in the bulk action (\ref{bulk action}) can be
written in terms of cosmological constant and massive potential
terms. Contraction of the field equation (\ref{field equations})
yields
\begin{equation}
R =\frac{2}{d_{2}}\left[ {\Lambda d + m_g^2{\cal X}}
\right],\,\,\,\,({\cal X} \equiv {g^{\mu \nu }}{{\cal X}_{\mu \nu
}}).
\end{equation}
The second term of the above equation can be summed with the
graviton's interaction Lagrangians, and then recast in a compact
form as
\begin{equation}
\frac{2}{{d_2}} m_g^2 {{\cal X} + m_g^2\sum\limits_{i = 1}^{d - 2}
{{c_i} {{\cal U}_i}} } = m_g^2 \sum\limits_{i = 1}^{d - 2} {(i -
2)\frac{{c_0^i{c_i}}}{{{r^i}}}} \prod\limits_{j = 3}^{i + 1}
{{d_j}},
\end{equation}
in which we have made use of the identity
\begin{equation}
2\prod\limits_{j = 3}^{i + 1} {{d_j}}  + \sum\limits_{y = 1}^i
{{{( - 1)}^y} \frac{{i!}}{{(i - y)!}}} \prod\limits_{j = 2}^{i - y
+ 1} {{d_j}}  = (i - 2)\prod\limits_{j = 3}^{i + 1} {{d_j}},
\end{equation}
where $\prod\limits_x^y {...}  = 1$ if $x>y$. As a result, the
on-shell action for the massive AdS BH  is
\begin{eqnarray} \label{BH on-shell action}
{{\cal I}_{BH}} &= &\frac{{\beta {\omega
_{{d_2}}^{(k)}}}}{{16\pi {G_d}}}\left[\frac{2r^{d_1}}{{{\ell ^2}}}{} -m_g^2\sum\limits_{i = 1}^{d - 2} {\frac{{(i - 2)c_0^i{c_i}{r^{d_{i +1}}}}}{{d - i - 1}}\prod\limits_{j = 3}^{i + 1} {{d_j}} } \right]_{{r_ +
}}^{R} \quad
\end{eqnarray}
where $R$ is an upper cutoff to regularize the on-shell actions.
Repeating the same procedure for the thermal AdS background
in massive gravity
yields
\begin{eqnarray} \label{AdS on-shell action}
 {{\cal I}_{AdS}} &= &\frac{{\beta_0 {\omega
_{{d_2}}^{(k)}}}}{{16\pi {G_d}}}\left[\frac{2R^{d_1}}{{{\ell ^2}}}{} -m_g^2\sum\limits_{i = 1}^{d - 2} {\frac{{(i - 2)c_0^i{c_i}{R^{d_{i +1}}}}}{{d - i - 1}}\prod\limits_{j = 3}^{i + 1} {{d_j}} } \right] \qquad
\end{eqnarray}
Fixing the temperature of both the AdS and the BH
configurations at $r=R$, i.e., ${\beta _0}V_0(R)^{1/2} = \beta
V{(R)^{1/2}}$ so that both the AdS and the BH
spacetimes must have the same geometry at $r=R$ gives ${\beta _0} = \beta \Big( {1 - \frac{{m{\ell ^2}}}{{2{R^{d - 1}}}}
+ O({R^{ - 2(d - 1)}})} \Big)$. Subtracting the
on-shell action of the AdS background from the AdS BH one
we find
\begin{eqnarray} \label{finite on-shell action}
{{\cal I}} &\equiv& \mathop {\lim }\limits_{R \to \infty } \left( {{{\cal I}_{BH}} - {{\cal I}_{AdS}}} \right) \nonumber \\
&=& \frac{{\beta {{\omega_{{d_2}}^{(k)}}}r_ + ^{{d_3}}}}{{16\pi {G_d}}}\Bigg[ {k - \frac{{r_ + ^2}}{{{\ell ^2}}} + m_g^2\sum\limits_{i = 1}^{d - 2} {\Big( {\frac{{(i - 1)c_0^i{c_i}}}{{r_ + ^{i - 2}}}\prod\limits_{j = 3}^i {{d_j}} } \Big)} } \Bigg] \nonumber\\
\end{eqnarray}
where the following identity has been used:
\begin{equation}
\frac{1}{{{d_2}}}\prod\limits_{j = 2}^i {{d_j}}  + \frac{{i -
2}}{{d - i - 1}}\prod\limits_{j = 3}^{i + 1} {{d_j}}  = (i -
1)\prod\limits_{j = 3}^i {{d_j}}.
\end{equation}


\section{Thermodynamics}

Thermodynamic quantities
associated with TBH spacetimes can be directly extracted via the
partition function,  Eqs. (\ref{partition function}) and
(\ref{finite on-shell action}). The mean energy of thermal
radiation, $\left\langle E \right\rangle$, is given by
\begin{eqnarray} \label{Mass}
M &\equiv& {\left\langle E \right\rangle}=  - \frac{\partial }{{\partial \beta }}\ln {\cal Z} \nonumber \\
&=& \frac{{{d_2}{\omega
_{{d_2}}^{(k)}}}}{{16\pi }}\Bigg[ {k + \frac{{r_{+}^2}}{{{\ell ^2}}} + m_g^2\sum\limits_{i = 1}^{d - 2} {\Big( {\frac{{c_0^i{c_i}}}{{{d_2}r_ + ^{i - 2}}}\prod\limits_{j = 2}^i {{d_j}} } \Big)} } \Bigg]r_ + ^{{d_3}}, \nonumber\\
\end{eqnarray}
which is in agreement with the ADM mass
\cite{AshtekarMagnon1984,AshtekarDas2000} of the BH
spacetime (setting $G_{d}=1$). Noting that the pressure is $P = - \Lambda /8\pi=
{d_1}{d_2}/16\pi \ell ^2$ (implying that the BH mass is
interpreted as the enthalpy, $M \equiv H$ \cite{Kastor2009}), we find
\begin{equation} \label{volume}
V = {\left( {\frac{{\partial M}}{{\partial P}}} \right)_{X_{i}}} =
\frac{{{\omega _{{d_2}}^{(k)}}}}{{{d_1}}}r_ + ^{{d_1}},
\end{equation}
for the thermodynamic volume, where $X_{i}$ denotes the extensive quantities. The Gibbs free
energy, which depends on the quantities $T$ and $P$, is given by
\begin{eqnarray} \label{Gibbs}
G &\equiv& -{\beta ^{ - 1}}\ln {\cal Z}(\beta ,P) \nonumber \\
&=&\frac{{{\omega_{{d_2}}^{(k)}}r_ + ^{{d_3}}}}{{16\pi }}\Bigg[ {k - \frac{{16\pi Pr_ + ^2}}{{{d_1}{d_2}}} + m_g^2\sum\limits_{i = 1}^{d - 2} {\Big( {\frac{{(i - 1)c_0^i{c_i}}}{{r_ + ^{i - 2}}}\prod\limits_{j = 3}^i {{d_j}} } \Big)} } \Bigg]. \nonumber\\
\end{eqnarray}
Finally, the entropy of TBHs is calculated as
\begin{equation} \label{entropy}
S  = \beta M  - {\cal I} = \frac{{{\omega _{{d_2}}^{(k)}}}}{{4}}r_
+ ^{{d_2}}.
\end{equation}

These quantities obey the Smarr formula
\begin{equation} \label{Smarr relation}
(d - 3)M = (d - 2)TS - 2PV + \sum\limits_{i = 1}^{d - 2} {(i -
2){{\cal C}_i}{c_i}},
\end{equation}
with ${{\cal C}_i} = {\left( {\frac{{\partial M}}{{\partial {c_i}}}}
    \right)_{S,P,{c_{i \ne j}}}} = \frac{{m_g^2{\omega
    _{{d_2}}^{(k)}}}}{{16\pi }}c_0^ir_ + ^{{d-i-1}}\prod\limits_{j =
    2}^i {{d_j}}$.

Note that Eq. (\ref{Smarr relation}) (which follows from Eulerian
scaling \cite{Kastor2009}), proves that variations in the
massive couplings ($c_{i}$) are required for consistency of the
extended first law of thermodynamics with the Smarr formula, so
massive couplings are not fixed \textit{a priori}. This implies that the
thermodynamic quantities ($M$, $T$, $S$, $P$, $V$, ${{\cal C}_i}$
and $c_{i}$) satisfy analytically the first law of thermodynamics
in the enthalpy representation, i.e. $dM = TdS + VdP +
\sum\limits_{i = 1}^{d - 2} {{{\cal C}_i}d{c_i}}$.


\section{Equation of state and phase structure}

The equation of state (EOS) is simply obtained from
(\ref{temperature}) as
\begin{equation} \label{pressure}
P = \frac{{{d_2}T}}{{4{r_ + }}} - \frac{{{d_2}{d_3}k}} {{16\pi r_
+ ^2}} - \frac{{m_g^2}}{{16\pi }}\sum\limits_{i = 1}^{d - 2}
{\Big( {\frac{{c_0^i{c_i}}}{{\,r_ + ^i}}\prod\limits_{j = 2}^{i +
1} {{d_j}} } \Big)}.
\end{equation}
The critical point occurs at the spike like divergence of specific
heat at constant pressure (i.e., an infection point in the $P-V$
diagrams) and can be found from the relations
\begin{equation} \label{critical point}
{\left( {\frac{{\partial P}}{{\partial v}}} \right)_{T}} = {\left(
{\frac{{{\partial ^2}P}}{{\partial {v^2}}}} \right)_{T}} = 0 \to
{\left( {\frac{{\partial P}}{{\partial {r_ + }}}} \right)_{T}} =
{\left( {\frac{{{\partial ^2}P}}{{\partial r_ + ^2}}} \right)_{T}}
= 0,
\end{equation}
which yield
\begin{equation} \label{critical point equation}
{2k + m_g^2\sum\limits_{i = 1}^{d - 2} {\Big( {\frac{{i(i -
1)c_0^i{c_i}}}{{r_ + ^{i - 2}}}\prod\limits_{j = 4}^{i + 1}
{{d_j}} } \Big)}  = 0},
\end{equation}
where the specific volume  $v  = 4{r_ + }\ell
_{\rm{P}}^{{d_2}}/{d_2}$ is proportional to $ r_ + $
\cite{KubiznakMann2012,GunasekaranEtal:2012}.

Since the expression on the left-hand side of Eq.
(\ref{critical point equation}) is a real and inhomogeneous
polynomial equation of degree ($d-4$) in $r_+$, an arbitrary
number of critical points could be produced by adjusting the
spacetime dimension $d$. Consequently higher dimensional TBHs
within the framework of massive gravity can exhibit as many as
$n=(d-4)$ critical points in $d$-dimensions. Note that criticality
does not take place in $d=4$; the inclusion of
$U(1)$ charge can yield critical behavior in this case.
Interestingly, in $d \ge 7$ dimensions, with up to five graviton
self-interaction potentials (${\cal U}_1$, ..., ${\cal U}_5$), we have
\begin{eqnarray} \label{triple point equation}
&&(k + m_g^2c_0^2{c_2})r_ +^3 + 3{d_4}m_g^2c_0^3{c_3}r_ + ^2 + 6{d_4}{d_5}m_g^2c_0^4{c_4}{r_ + } \nonumber \\
&&+ 10{d_4}{d_5}{d_6}m_g^2c_0^5{c_5} = 0,
\end{eqnarray}
which can have three positive roots, indicating the existence of a
triple point and an analogue of the solid/liquid/gas phase
transition for uncharged-AdS BHs in massive gravity
(without loss of generality, we set $c_{0}=1$ hereafter). This
situation is similar to that seen in multi-spinning ($d\geq 6$)
Kerr-AdS BHs \cite{Altamirano2014}. In a $d$-dimensional
spacetime, an inhomogeneous polynomial equation with degree of
$(d-4)$ could at most have $(d-4)$ positive roots. This indicates
the possibility of finding more than three different BH
phases in massive gravity.

\section{Triple point and SBH/IBH/LBH phase transition}

\begin{figure}[tbp]
\epsfxsize=9cm \epsffile{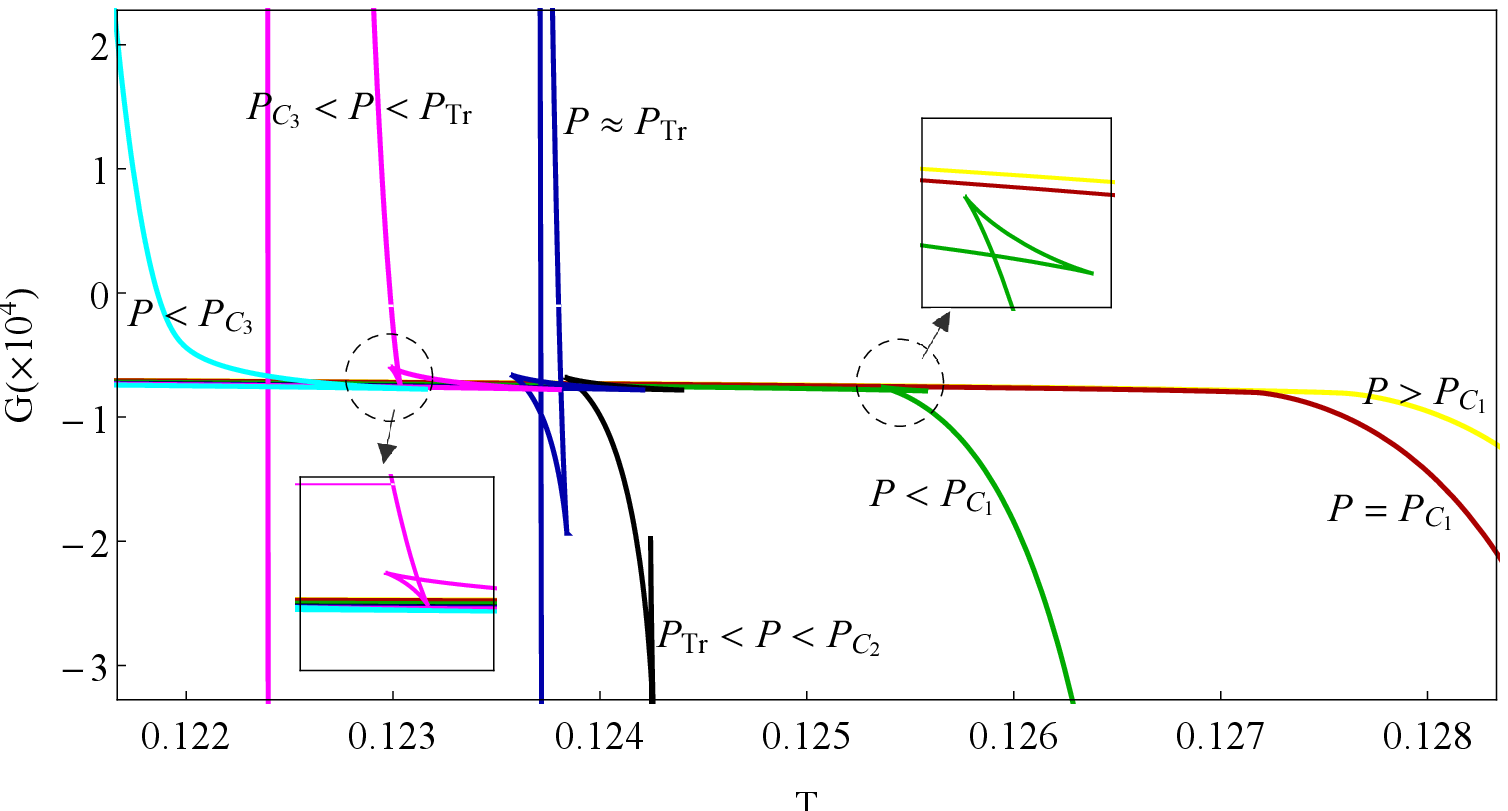}
    \caption{Analogue of a triple point: $G-T$ diagram for various pressures; we have set $k=0$, $d=10$, $m_{g}=1$, $c_{0}=1$, $c_{1}=1$, $c_{2}=1.6$, $c_{3}=-2.1$, $c_{4}=0.35$, $c_{5}=2.6$, $c_{6}=-2.9$, $c_{7}=0$, and $c_{8}=0$.\\
        \textit{Critical data}: ($T_{C_{1}}=0.127185$, $P_{C_{1}}=0.002482$), ($T_{C_{2}}=0.124382$, $P_{C_{2}}=0.001624$), and ($T_{Tr}=0.123577$, $P_{Tr}=0.001532$)}
    \label{GT3}
\end{figure}

We now consider a ten-dimensional BH spacetime with a
flat ($k=0$) geometry for its event horizon. This leads to a
boundary dual gauge theory with a Minkowski metric. To
observe the analogue of a triple point, we have tuned the
massive couplings to produce three critical points, as shown
in Fig. \ref{GT3}, where we depict the $G-T$  diagram corresponding
to the SBH/IBH/LBH phase transition that resembles
the solid/liquid/gas phase transition in usual substances.
The isobar corresponding to $P_{C_{2}}<P<P_{C_{1}}$
displays the
swallowtail (vdW) behavior which indicates a first-order
phase transition. For pressures with $P_{Tr}<P<P_{C_{2}}$ 
we
observe two swallowtails, indicating the appearance of
two first-order phase transitions, implying three-phase
behavior. By decreasing the pressure, the two swallowtails
eventually merge and a triple point ($T_{Tr}$,$P_{Tr}$) appears.

\section{\textit{N}-fold RPT$\rm{\textbf{s}}$}

\begin{figure}[tbp]
	\epsfxsize=9cm \epsffile{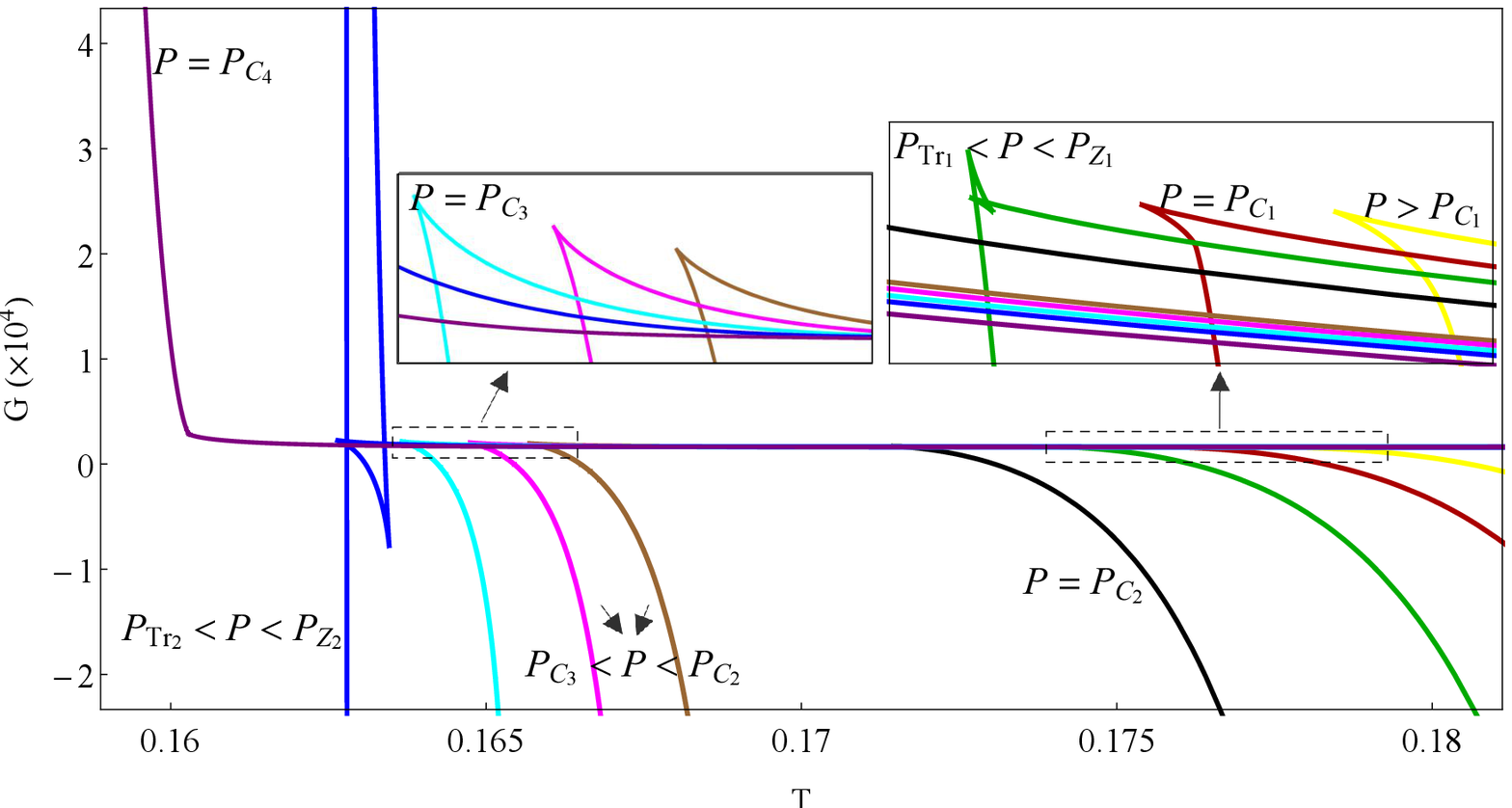}
	\caption{Twofold RPTs: $G-T$ diagram for various pressures; we have set $k=0$, $d=10$, $m_{g}=1$, $c_{0}=1$, $c_{1}=1$, $c_{2}=3$, $c_{3}=-4.3$, $c_{4}=2.6$, $c_{5}=1.2$, $c_{6}=-3.5$, $c_{7}=2.6$, and $c_{8}=0$.\\
		\textit{Critical data}: ($T_{C_{1}}=0.175778$, $P_{C_{1}}=0.007513$), ($T_{C_{2}}=0.172218$, $P_{C_{2}}=0.005847$), ($T_{C_{3}}=0.1653102$, $P_{C_{3}}=0.003276$), ($T_{C_{4}}=0.160285$, $P_{C_{4}}=0.002343$), ($T_{Tr_{1}}=0.174034$, $P_{Tr_{1}}=0.006794$), ($T_{Z_{1}}=0.174203$, $P_{Z_{1}}=0.006882$), ($T_{Tr_{2}}=0.162519$, $P_{Tr_{2}}=0.0029496$), and ($T_{Z_{2}}=0.162572$, $P_{Z_{2}}=0.002965$)}
	\label{GT4}
\end{figure}

\begin{figure}[tbp]
	\epsfxsize=8.5cm \epsffile{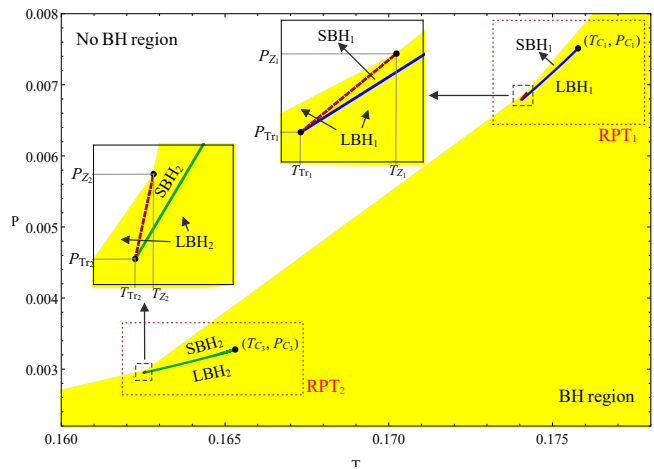}
	\caption{Phase diagram for twofold RPTs: The corresponding coexistence lines of twofold RPTs, presented in Fig. \ref{GT4}, in the $P-T$ diagram. Dashed and solid lines represent zeroth- and first-order phase transitions, respectively. Qualitatively, this behavior is generic for any twofold RPTs.}
	\label{PT}
\end{figure}

Next we present an analogue of four critical points in the context of BH thermodynamics. By further tuning the parameters, we show in Fig. \ref{GT4} that four critical points can
be created in the thermodynamic phase space of ten-dimensional
planar BHs. As seen in Fig. \ref{GT4}, two distinct
reentrant phase transitions take place in the phase space,
indicating the appearance of two virtual triple points
(referred to as $P_{Tr_{1}}$ and $P_{Tr_{2}}$). In fact, for a fixed pressure
in ranges of $P_{Tr_{1}}<P<P_{Z_{1}}$ and
$P_{Tr_{2}}<P<P_{Z_{2}}$, 
, each
RPT takes place along a single horizontal transition line in
the $P-T$ diagram (Fig.
\ref{PT}) and we may observe multiple
phase transitions indicating $N$-fold reentrant phase transition
behavior. In these two distinct regions of phase space,
as temperature decreases monotonically, a first-order phase
transition is observed, and then, a finite jump (discontinuity)
appears in the global minimum of the Gibbs free energy, which displays the zeroth-order phase transition.
Then we observe a twofold reentrant phenomenon, which
in some aspects is reminiscent of the sequence of reentrant
phase transitions in liquid crystals \cite{MultipleRPTs1986,MultipleRPTs1988Review,MultipleRPTs2002Book}. But an important
difference remains here. In the thermodynamic phase
space of the obtained TBHs, two RPTs are observed in
distinct regions of phase space, i.e., the first RPT occurs for
a given pressure in $P_{Tr_{1}}<P<P_{Z_{1}}$
as temperature is
lowered monotonically in the region $T_{Tr_{1}}<T<T_{Z_{1}}$,
and the second RPT occurs in the same way with temperature
in the region $T_{Tr_{2}}<T<T_{Z_{2}}$ and a given pressure in
$P_{Tr_{2}}<P<P_{Z_{2}}$. But, in liquid crystals and multicomponent
fluids, multiple RPTs (especially the double
reentrance sequence $A\to B \to A \to B$
\cite{DoubleReentrance1980,DoubleReentrance1988,DoubleReentrance1989,DoubleReentrance2003}) take place
as the temperature is lowered with other thermodynamic
quantities (such as pressure) held fixed. This situation is
more similar to that seen in ($d=7$) Lovelock-AdS BHs
with hyperbolic horizon topology \cite{Frassino2014}, in which a double
RPT (LBH $\to$ SBH $\to$ LBH $\to$ SBH) has been observed
solely by lowering the temperature. These remarks indicate
the importance of studying the critical behavior of TBHs
in the other gravitational alternatives such as Lovelock gravity.

Summarizing, we have demonstrated the existence of
$N$-fold RPTs within massive gravity. These depend on the
number of inflection points. Our investigations show that
the existence of $N$-fold RPTs with corresponding ($N$)
virtual triple points is a generic feature of all types of
TBHs in higher-dimensional massive gravity, which is a
step forward in understanding BH phase transitions via
modified gravity. Perhaps these new kinds of $N$-fold RPTs
are present in many-body systems; they certainly merit
further exploration.

\section{Closing Remarks}

Since condensed matter systems usually break translational
invariance in nature, we have considered holographic
duals of such systems with homogeneity and isotropy
properties using the language of dRGT massive gravity
with a singular, degenerate reference metric. We obtained
TBH solutions that are free from pathological behavior
and dual to matter with broken translational symmetry
property; so, in principle, they can be dual to liquid crystals
\cite{Science2011,Beekman2017PhysicsReports,Beekman2017PRB}. These TBHs yield up to $n=(d-4)$ critical
points in $d$ dimensions. Consequently, a number of new
interesting phenomena emerge, in particular $N$-fold reentrant
phase transitions, indicating multiple phases in
$d \ge 7$ dimensions. Moreover, in the grand canonical
ensemble of charged-AdS BHs, this behavior persists
and holds in higher dimensions as well. Since (multiple)
RPTs are typical of liquid crystals, elementally, holographic
models for them using BHC could probably
simulate realistic critical behavior and perhaps be implemented
to predict new features of criticality, such as $N$-fold
RPTs, in nature. No proposal has yet been provided to
experimentally verify TBH phase transitions, but it is
conceivable that analogue gravity simulations \cite{AnalogueGravity2011Review}
one day get to this point.

From the molecular point of view, (multiple) RPTs
appear in compounds (especially in liquid crystals) and
only approximate qualitative explanations exist for them.
Regarding this, we believe that the analytic TBH equations
of state (\ref{pressure}) in higher dimensions and their counterparts in other
gravitational alternatives such as Lovelock gravity \cite{Frassino2014}
could possibly shed some light on the microscopic structure
of multiple RPTs (or perhaps $N$-fold RPTs if they exist) in
liquid crystals and multicomponent fluids \cite{MultipleRPTs1986,MultipleRPTs1988Review,MultipleRPTs2002Book} which merits further investigation in the future. Furthermore, we
expect that the exact massive TBHs with a range of novel
phase transitions constructed here have potential applications
in the context of the AdS/CFT correspondence; in
particular, a holographic interpretation of them remains an
open question.

Finally, we note that our considerations are valid for all
types of TBHs, which can be manifest by introducing the
effective topological factor $k_{\rm{eff}}=(k + m_g^2c_0^2{c_2})$ appearing
in Eqs. (\ref{metric function}), (\ref{temperature}),
(\ref{Gibbs}), (\ref{pressure}), and (\ref{critical point equation}). The only necessity is
that the same value must be provided for the combination
$k_{\rm{eff}}$ (by varying the massive constant c2) while keeping
other parameters fixed. As a result, the same critical points with the same critical behavior are found for the cases of the
spherical, planar, and hyperbolic BHs.


\begin{acknowledgements}
We would like to thank Andrew Tolley for helpful
correspondence, and anonymous referees whose comments
helped improve our paper. AD and SHH wish to thank
Shiraz University Research Council. AD would like to
thank Soodeh Zarepour for providing \textit{Mathematica} programming
codes and numerous stimulating discussions. The work of RBM was supported by the Natural
Sciences and Engineering Research Council of Canada.
\end{acknowledgements}


\begin{thebibliography}{99}

    \bibitem{deRham2010Gabadadze}C. de Rham and G. Gabadadze, Phys. Rev. D \textbf{82}, 044020 (2010).

    \bibitem{dRGT}C. de Rham, G. Gabadadze and A.J. Tolley, Phys. Rev. Lett. \textbf{106}, 231101 (2011).

    \bibitem{MassiveCosmologies2011}G. D'Amico, et al., Phys. Rev. D \textbf{84},124046 (2011).

    \bibitem{Akrami2013Koivisto}Y. Akrami, T.S. Koivisto and M. Sandstad, JHEP \textbf{03}, 99 (2013).

    \bibitem{Akrami2015Hassan}Y. Akrami, S.F. Hassan, F. Knnig, A. Schmidt-May, and A.R. Solomon, Phys. Lett. B \textbf{748}, 37 (2015).

    \bibitem{Schmidt-May2016JCAP}E. Babichev, et al., JCAP \textbf{09}, 016 (2016).

    \bibitem{LIGO2017} LIGO scientific collaboration, Phys. Rev. Lett. \textbf{118}, 221101 (2017).

    \bibitem{deRhamREVIEW2014}C. de Rham, Living Rev. Rel. \textbf{17}, 7 (2014).

    \bibitem{BDghost1972}D.G. Boulware and S. Deser, Phys. Rev. D \textbf{6}, 3368 (1972).

    \bibitem{HassanRosen2012PRL}S.F. Hassan and R.A. Rosen, Phys. Rev. Lett. \textbf{108}, 041101 (2012).

    \bibitem{HassanRosenSchmidt-May2012JHEP} S.F. Hassan, R.A. Rosen, and A. Schmidt-May, JHEP \textbf{02}, 026 (2012).

    \bibitem{TQDO2016a}T.Q. Do, Phys. Rev. D \textbf{93}, 104003 (2016).

    \bibitem{TQDO2016b}T.Q. Do, Phys. Rev. D \textbf{94},044022 (2016).

    \bibitem{Vegh2013} D. Vegh, arXiv:1301.0537

    \bibitem{BlakeTong2013} M. Blake and D. Tong, Phys. Rev. D \textbf{88}, 106004 (2013).

     \bibitem{Alberte2016}L. Alberte and A. Khmelnitsky, , Phys. Rev. D \textbf{91}, 046006 (2015)

    \bibitem{Alberte2016SolidHolography} L. Alberte, M. Baggioli, A. Khmelnitsky and O. Pujolas, JHEP \textbf{02}, 114 (2016).

    \bibitem{Alberte2018HolographicPhonons} L. Alberte,M. Ammon, A. Jiménez-Alba, M. Baggioli and Oriol Pujolas, Phys. Rev. Lett. \textbf{120}, 171602 (2018).

    \bibitem{Alberte2018BHelasticity} L. Alberte, M. Ammon, M. Baggioli, A. Jiménez and O. Pujolàs, JHEP \textbf{01}, 129 (2018).

    \bibitem{CQG2017Review} D. Kubiznak, R.B. Mann, and M. Teo, Class. Quant. Grav. \textbf{34}, 063001 (2017).

    \bibitem{PVmassive2015}J. Xu, L.M. Cao, and Y.P. Hu, Phys. Rev. D \textbf{91}, 124033 (2015).

    \bibitem{TBH2017MassiveGravity} S.H. Hendi, , R.B. Mann, S. Panahiyan, and B. Eslam Panah, Phys. Rev. D \textbf{95}, 021501(R) (2017).

    \bibitem{ReentrantMassive2017}D. Zou, R. Yue, and M. Zhang, Eur. Phys. J. C \textbf{77}, 256 (2017).

    \bibitem{Frassino2014} A.M. Frassino, D. Kubiznak, R.B. Mann and F. Simovic, JHEP \textbf{09}, 080 (2014).

     \bibitem{DoubleReentrance1980}F. Hardouin and A.M. Levelut, J. Phys. (Paris) \textbf{41}, 41(1980).

     \bibitem{DoubleReentrance1988}V. N. Raja, R. Shashidhar, B. R. Ratna, G. Heppke and Ch. Bahr, Phys. Rev. A \textbf{37}, 303 (1988).

     \bibitem{DoubleReentrance1989}K. Ema, G. Nounesis, C.W. Garland and R. Shashidhar, Phys. Rev. A \textbf{39}, 2599 (1989).

     \bibitem{DoubleReentrance2003}X.F Han, S.T. Wang, A. Cady, M.D. Radcliffe and C.C. Huang, Phys. Rev. Lett. \textbf{91}, 045501 (2003).

    \bibitem{MultipleRPTs1986} J.O. Indekeu and A. Nihat Berker, Physica A: Statistical Mechanics and its Applications \textbf{140}, 368 (1986).

    \bibitem{MultipleRPTs1988Review}P.E. Cladis, Molecular Crystals and Liquid Crystals \textbf{165}, 85 (1988).

    \bibitem{MultipleRPTs2002Book}S. Singh and D.A. Dunmur, Liquid crystals: fundamentals, World Scientific, 2002.

    \bibitem{NarayananKumar:1994} T. Narayanan and A. Kumar, Phys. Rep. \textbf{249}, 135  (1994).

    \bibitem{Science2011}A. Mesaros, et al., Science \textbf{333}, 426 (2011).

   \bibitem{Alberte2013}L. Alberte and A. Khmelnitsky, Phys. Rev. D \textbf{88}, 064053 (2013).


    \bibitem{Cai2015}R.G. Cai, Y.P. Hu, Q.Y. Pan, and Y.L. Zhang, Phys. Rev. D. \textbf{91}, 024032 (2015).
    
    \bibitem{Mann:1997iz} 
    R.~B.~Mann,
    Annals Israel Phys.\ Soc.\  {\bf 13}, 311 (1997)
    [gr-qc/9709039].

    \bibitem{HawkingPage1983}S. Hawking and D. Page, Commun. Math. Phys. \textbf{87}, 577 (1983).

    \bibitem{Witten1998b} E. Witten, Adv. Theor. Math. Phys. \textbf{2}, 505 (1998).

    \bibitem{AshtekarMagnon1984}A. Ashtekar and A. Magnon, Class. Quant. Grav. \textbf{1}, L39 (1984).

    \bibitem{AshtekarDas2000}A. Ashtekar and S. Das, Class. Quant. Grav. \textbf{17}, L17 (2000).

    \bibitem{Kastor2009}D. Kastor, S. Ray, and J. Traschen, Class. Quant. Grav. \textbf{26}, 195011 (2009).

    \bibitem{KubiznakMann2012}D. Kubiznak and R. B. Mann, JHEP \textbf{07}, 033 (2012).

    \bibitem{GunasekaranEtal:2012} S. Gunasekaran, D. Kubiznak, and R.B. Mann, JHEP {\bf 11}, 110 (2012).

    \bibitem{Altamirano2014}N. Altamirano, D. Kubiznak, R. B. Mann, and Z. Sherkatghanad, Class. Quant. Grav. \textbf{31}, 042001 (2014).

    \bibitem{Beekman2017PhysicsReports} A.J. Beekman, et al., Phys. Rep. \textbf{683}, 1 (2017)

    \bibitem{Beekman2017PRB} A.J. Beekman, J. Nissinen, K. Wu, and J. Zaanen, Phys. Rev. B \textbf{96} 165115, (2017)

    \bibitem{AnalogueGravity2011Review} C. Barcelo, S. Liberati, and M. Visser, Living Rev. Rel. \textbf{14}, 3 (2011).

\end{thebibliography}
\end{document}